# Decoding Knowledge Claims: The Evaluation of Scientific Publication Contributions through Semantic Analysis

Luca D'Aniello[*], Nicolás Robinson-García[**], Massimo Aria[***], and Corrado Cuccurullo[****]

[*]*luca.daniello@unina.it*
ORCID: 0000-0003-1019-9212
Department of Social Sciences, K-Synth Academic spin-off, University of Naples Federico II, Italy

[**] *elrobin@ugr.es*
ORCID: 0000-0002-0585-7359
Unit for Computational Humanities and Social Sciences (U-CHASS), EC3 Research Group, University of Granada, Spain

[***] *massimo.aria@unina.it*
ORCID: 0000-0002-8517-9411
Department of Economics and Statistics, K-Synth Academic spin-off, University of Naples Federico II, Italy

[****] *corrado.cuccurullo@unicampania.it*
ORCID: 0000-0002-7401-8575
Department of Management, University of Campania Luigi Vanvitelli, Italy
Department of Economics and Statistics, K-Synth Academic spin-off, University of Naples Federico II, Italy

The surge in scientific publications challenges the use of publication counts as a measure of scientific progress, requiring alternative metrics that emphasize the quality and novelty of scientific contributions rather than sheer quantity. This paper proposes the use of Relaxed Word Mover's Distance (RWMD), a semantic text similarity measure, to evaluate the novelty of scientific papers. We hypothesize that RWMD can more effectively gauge the growth of scientific knowledge. To test such an assumption, we apply RWMD to evaluate seminal papers, with Hirsch's H-Index paper as a primary case study. We compare RWMD results across three groups: 1) H-Index-related papers, 2) scientometric studies, and 3) unrelated papers, aiming to discern redundant literature and hype from genuine innovations. Findings suggest that emphasizing knowledge claims offers a deeper insight into scientific contributions, marking RWMD as a promising alternative method to traditional citation metrics, thus better tracking significant scientific breakthroughs.

## 1. Introduction

The growth of scientific knowledge has traditionally been measured by the volume of publications (Price, 1963) —a metric that has served as the primary gauge of productivity and impact in academia for decades. This quantitative approach has long been used in the evaluation of academic progress and the allocation of research funding. However, the reliability of publication counts as a metric of scientific advancement is becoming increasingly questioned (Chen, 2012). In recent years, the academic world has been inundated by an unprecedented number of published papers (Nane et al., 2023), a phenomenon that has led to concerns about the saturation of literature and the genuine contribution of many works (Landhuis, 2016). This growth in publication volume has raised significant doubts about the validity and value of contributions, leading to what can be described as an information overproduction (Arnold et al., 2023). In response to these challenges, there is a growing consensus on the need for alternative metrics that can more accurately identify true innovation and contributions of papers (Shibayama and Wang, 2019). Several approaches were proposed for redefining how scientific knowledge is measured (Wang, 2016; Azoulay et al., 2011; Foster et al., 2015; Shibayama & Wang, 2019). These methodologies focus on the analysis of knowledge claims, which are defined as novel and original contributions to the current state-of-the-art within specific fields (Shibayama and Wang, 2019). Still, they are all based on citation and publication metrics.

Accordingly, Natural Language Processing (NLP) techniques can help to identify knowledge claims, representing granular, qualitative approaches to assessing the content of scientific publications, focusing on the substantive quality and novelty of the research rather than merely its presence in the literature.

This paper proposes an innovative NLP approach to evaluate these knowledge claims. By shifting the focus from sheer output to the quality of contributions, this method aims to provide a more nuanced understanding of scientific growth, reducing the emphasis on publication quantity as the sole indicator of scientific progress. Through the application of semantic statistical analysis, we aim to explore the textual similarities and differences within scientific papers, thereby identifying not only redundancies and hype but also truly disruptive innovations that push the boundaries of knowledge.

## 2. Methodology

The development of effective text similarity measurement techniques is central to numerous applications in NLP. Traditional approaches often rely on measuring the overlap of terms between documents. These methods, such as Bag-of-Words and TF-IDF, focus on quantifying the direct textual overlap without considering the semantic relationships between words that appear in different texts (Qader et al., 2019; Dumais et al., 1988; Blej et al., 2003). Such approaches are intuitive and straightforward but fall short of capturing the nuanced semantic similarities that might exist between texts with little to no direct term overlap. This is a significant drawback in many real-world scenarios where the context and semantic richness of language play a crucial role in understanding the similarity between documents. To address the semantic gap, the field of NLP has turned towards word embeddings. Word embeddings represent words in a high-dimensional space as vectors. These vectors capture semantic properties such that semantically similar words are located in proximity within this space. Models like Word2Vec, GloVe, and FastText are popular examples of word embedding (Mikolov et al., 2013a; Mikolov et al., 2013b; Pennington et al., 2014).

An innovative approach that leverages the power of word embeddings to measure text similarity is the Word Mover's Distance (WMD), introduced by Kusner et al. (2015). Based on the classic Earth Mover's Distance (EMD), WMD calculates the minimum cumulative distance for words from one document to exactly match the word distribution of another document in the embedding space. The goal is to compute the minimum cost of transforming one document representation into another, using the distance between embeddings as the cost of word transformation. This method effectively captures the semantic differences and similarities between texts, even if they have no words in common. While effective, the computational demands of WMD pose challenges when applied to large documents or those with a high number of unique words. The computational cost arises from solving the transportation problem, which is notably high, requiring super cubic time (Atasu et al., 2017; Wu et al., 2018). To mitigate this, the Relaxed Word Mover's Distance (RWMD) was proposed (Atasu, 2017), simplifying the optimization problem of WMD, and making it more computationally efficient while still leveraging the semantic richness of word embeddings. This is achieved by relaxing a constraint in the original WMD formulation, which reduces the computational complexity from super-cubic to linear time, depending on the implementation. As a result, RWMD allows for the rapid computation of semantic text similarity across large datasets, making it a practical choice for real-world applications where speed and scalability are crucial. Moreover, RWMD, given its simplicity and speed, emerges as one of the most successful outcomes of this research endeavor. In literature, some studies demonstrate its competitive quality compared to WMD, with the added benefit of significantly faster processing, especially for extensive vocabularies (Atasu et al., 2017; Werner & Laber, 2019).

## 3. Case study

To assess the effectiveness of RWMD in identifying knowledge redundancies or claims in scientific papers, we focus on the seminal paper by Hirsch (2005) that introduced the H-Index, a metric that has been widely discussed and critiqued within the academic community. The rapid adoption and extensive application of the H-Index resemble a hype rather than a reflection of substantial intellectual advancement (Rousseau et al., 2013). To investigate these claims, we employ the RWMD to analyze the semantic similarity between Hirsch's paper and other scientific papers across different categories. In particular, we manually extract the sections of papers relative to the authors' contributions, ideas, and proposals. Given the typical structure of scientific papers (e.g., IMRaD), we retrieved the conclusion sections. However, the Hirsch document lacked distinct sections. Therefore, two selectors (LDA and MA) identified sentences containing the manuscript's proposal and novelty. Then, we measured the similarity distance of the H-Index document with three distinct groups:

- **H-Index papers**: Papers that discuss or analyze the H-Index, derived from the review by Alonso et al. (2009). The comparison between Hirsch's original H-Index paper and subsequent papers that focus specifically on the H-Index allows for a direct evaluation of the influence and evolution of the H-Index within its own domain. This group is crucial for understanding how Hirsch's ideas have been expanded upon, critiqued, or reinforced over time. By analyzing these papers, we can assess the degree of intellectual engagement with the H-Index concept and identify whether it has fostered incremental innovations or simply repeated the original ideas without significant additions.
- **Scientometrics papers**: random papers published on Scientometrics that do not focus on the H-Index. This comparison aims to highlight thematic divergences or convergences within the field that may not be directly influenced by the H-Index. The papers were retrieved from Scopus by searching all publications in Scientometrics journal without any reference to the H-index or Hirsch.
- **Random papers**: A randomly selected set of papers from various disciplines to gauge the H-Index paper's influence or mention outside its immediate field, helping identify any interdisciplinary impact or lack thereof.

These groups are selected with the expectation of computing RWMD scores which will be likely to vary significantly, reflecting the contextual usage of the H-Index across disciplines. Some fields may show unexpectedly lower scores due to the adoption of the H-Index as a performance metric, while others may show no significant semantic closeness, indicating limited or no influence. By clearly delineating these groups and articulating the specific expectations of each, the study enhances the robustness of its methodological approach.

The RWMD was employed to measure the semantic similarity between the Hirsch paper, labeled as the query document, and the papers in each group. A total of 32 papers related to the H-Index were retrieved from Alonso et al.'s review (2009), and subsequently, 32 papers were collected for both the Scientometrics and Random groups to match this number. The word embedding used for the RWMD was GloVe (Pennington et al., 2014), an unsupervised model for word vector representation. GloVe is a count-based model that learns word vectors from their co-occurrence information, determining how frequently words appear together in large text corpora. It is a fast word embedding technique trained on word-word co-occurrence probabilities, capable of capturing semantic meanings that can be expressed as vector differences (Mohammed et al., 2020). Before employing the RWMD, the texts were preprocessed, tokenizing the texts, converting all uppercase letters to lowercase, removing extra spaces, punctuation, and stopwords. Following preprocessing, the GloVe model embedding was employed, assigning a vector to each word of the texts and representing them in an embedding vector space. Proximity in this space indicates similarity between words. Cosine similarity was used to measure the distance between word vectors, quantifying the cosine of the angle between them, with values ranging from 0 to 1. A cosine similarity of 1 signifies identical words, while 0 indicates no similarity.

Finally, we computed the RWMD between the query document (Hirsch's H-Index paper) and the documents from three groups. It measures the cost of transforming words from the query document into words from each document in the collection. For each document, RWMD was calculated relative to the query document, obtaining a value between 0 and 1, where 0 represents complete dissimilarity and 1 signifies semantic identical texts. Additionally, a Kruskal-Wallis test was performed among the distance values of the three groups. Subsequently, pairwise comparisons were conducted using the Wilcoxon rank sum exact test to determine the statistical significance of differences between distances among pairs of groups.

By employing the aforementioned semantic statistical-based approach, we hypothesize that if there is indeed a hype, the distance of H-indices from the query document would be smaller compared to that of the Scientometrics and Random groups.

All analyses were conducted in R (v. 4.3.1). The RWMD distance and the GloVe model were computed using the *text2vec* package. Data including the extracted texts from all selected articles is freely available at the following link: https://zenodo.org/records/10997154.

**4. Findings**

The results of the RWMD are presented in Table 1, where the semantic similarity distance between each paper in the groups and the query document was measured. For each group, the median distance values and interquartile ranges are reported. Our findings confirm our hypothesis, demonstrating that the H-Index papers show a higher level of similarity with the query document (median = 0.5763) compared to papers in the second and third groups (medians= 0.4331 and 0.3466, respectively). Additionally, the similarity distances among groups were statistically significant, and pairwise comparisons using the Wilcoxon rank sum exact test also yielded statistical significance across all pairs of groups.

Using RWMD, we effectively assessed the text similarity concerning the Hirsch documents. Specifically, focusing on the H-Index papers, RWMD successfully identified related papers with minimal or broader contributions. For instance, the top three documents include a study on the introduction of the h(2)-index as an alternative to the h-index for evaluating scientific output (document 7), a paper on the use of the g-index to enhance the h-index for assessing scientific impact (document 1), and a study on the introduction of three refinements of the $h_1$-index ($h_1^+$, $h_1^*$, $h_1^A$), as well as two refined h-indices ($h^x$ and $h^*$) (document 10). The RWMD findings not only aid in detecting papers with similar or minimal contributions but also in identifying papers proposing entirely novel or different concepts compared to the query document. This may signify the introduction of new knowledge or topics unrelated to the research area. Such papers are ranked lowest in our findings, with a low similarity distance. Specifically, focusing on the last two documents, document 24 stands out as the only paper in the groups discussing the widespread use of the H-index for evaluating individual scientists, while highlighting its biases and limitations. On the other hand, document 23 delves into the utilization of citation data to evaluate scientific authors, raising questions about the reliability of selected indicators and their effectiveness in distinguishing between authors.

In the Scientometrics group, the top three similar documents identified by the RWMD are a study related to interdisciplinary research spanning from 1975 to 2005 across six domains, utilizing established indicators and a new Diversity Index (document 6); research on the assessment of four topic modeling algorithms using a test dataset from seven scientific fields, comparing different algorithms and indices (document 15); and the description of indexes measuring collaborative strength in a discipline (document 28). Conversely, the last two documents focus on measures to quantify academic bibliographic database sizes, criticizing Google Scholar (document 8), and the use of word embedding in the mathematics research domain (document 25). Although all papers in this group are not related to the H-Index topic, it is interesting how the RWMD effectively captured semantic similarities, particularly evident in the contribution summarized in the text of the top three documents regarding the proposal of indexes.

Finally, the Random Papers group shows the lowest similarity distance in relation to the query document. Indeed, their contributions are entirely unrelated to the Hirsch proposal. For instance, the first document is a survey that evaluates current fake news research by defining and differentiating it from related concepts, detailing interdisciplinary research, and reviewing detection methods from multiple perspectives (document 7). Meanwhile, the last two documents focus on AI applications in the water domain and a medical study about cell segregation (documents 28 and 31, respectively).

Table 1. RWMD similarity distances between each document and the query document are presented. Median values and interquartile ranges are reported for each group.

| *H-Index papers* Doc ID - Distance | *Scientometrics papers* Doc ID - Distance | *Random papers* Doc ID - Distance |
|---|---|---|
| 7 - 0.7636 | 6 - 0.5435 | 7 - 0.5354 |
| 1 - 0.6729 | 15 - 0.5416 | 13 - 0.4784 |
| 10 - 0.6701 | 28 - 0.5367 | 21 - 0.4453 |
| 32 - 0.6680 | 1 - 0.5070 | 11 - 0.4329 |
| 2 - 0.6658 | 11 - 0.5030 | 2 - 0.4314 |
| 31 - 0.6583 | 31 - 0.4897 | 12 - 0.4261 |
| 20 - 0.6536 | 17 - 0.4881 | 8 - 0.4160 |
| 19 - 0.6276 | 20 - 0.4764 | 19 - 0.3992 |
| 11 - 0.6132 | 5 - 0.4760 | 18 - 0.3990 |
| 13 - 0.6130 | 2 - 0.4665 | 20 - 0.3837 |
| 8 - 0.6093 | 23 - 0.4654 | 25 - 0.3830 |
| 5 - 0.6065 | 27 - 0.4599 | 10 - 0.3798 |
| 14 - 0.5973 | 30 - 0.4479 | 24 - 0.3646 |
| 22 - 0.5935 | 22 - 0.4444 | 15 - 0.3594 |
| 3 - 0.5902 | 10 - 0.4443 | 3 - 0.3530 |
| 15 - 0.5788 | 13 - 0.4379 | 5 - 0.3487 |
| 9 - 0.5738 | 3 - 0.4283 | 23 – 0.3446 |
| 4 - 0.5678 | 29 - 0.4194 | 16 - 0.3277 |
| 28 - 0.5630 | 18 - 0.4174 | 9 - 0.3192 |
| 16 - 0.5575 | 9 - 0.4165 | 6 - 0.3172 |
| 17 - 0.5375 | 19 - 0.4100 | 32 - 0.3071 |
| 29 - 0.5372 | 14 - 0.4031 | 26 - 0.2973 |
| 25 - 0.5344 | 4 - 0.3977 | 30 - 0.2958 |
| 12 - 0.5210 | 12 - 0.3909 | 14 - 0.2893 |
| 18 - 0.5200 | 7 - 0.3904 | 1 - 0.2835 |
| 26 - 0.5138 | 32 - 0.3890 | 17 - 0.2821 |
| 21 - 0.5121 | 24 - 0.3811 | 29 - 0.2695 |
| 27 - 0.5053 | 16 - 0.3802 | 4 - 0.2573 |
| 6 - 0.4886 | 26 - 0.3628 | 27 - 0.2494 |
| 30 - 0.4652 | 21 - 0.3498 | 22 - 0.2481 |
| 24 - 0.4555 | 8 - 0.3230 | 28 - 0.2210 |
| 23 - 0.4265 | 25 - 0.3165 | 31 - 0.1862 |
| **Median** | **0.5763** | **0.4331** | **0.3466** |
| **[IQR]** | [0.5208-0.6131] | [0.3908-0.4689] | [0.2878-0.3875] |

Table 2. Kruskal-Wallis test to compare the distances among the three groups. followed by pairwise comparisons using the Wilcoxon rank sum exact test.
*Significant $0.01 < p \leq 0.05$, **Significant $p \leq 0.01$

|  | p-value |
|---|---|
| **Kruskall-Wallis test** | <0.001** |
| **Pairwise comparisons using Wilcoxon rank sum exact test** |  |
| H-Index papers vs Scientometrics papers | <0.001** |
| H-Index vs Random papers | <0.001** |
| Scientometrics papers vs Random papers | <0.001** |

## 5. Conclusion and further developments

This study demonstrates the utility of employing the RWMD to evaluate the semantic similarity of scientific papers within and across different domains. By focusing on the seminal Hirsch H-Index paper as a query document, our analysis has provided a structured approach to measuring the intellectual proximity and thematic relevance of subsequent publications within the fields of scientometrics and beyond.

Our findings confirm that the H-Index papers, as expected, show a higher level of similarity with the Hirsch paper, affirming the prevalent influence of the H-Index in discussions surrounding academic impact metrics. This underscores a saturation within the domain, where many papers tend to reiterate or slightly modify existing concepts rather than proposing groundbreaking new ideas. The significant similarity scores within this group point to a possible overemphasis on incremental advancements rather than substantial scholarly contributions.

Conversely, the analysis of papers from unrelated fields and other areas of scientometrics using RWMD effectively highlighted a broader range of thematic engagements. Specifically, the lower similarity scores for these papers indicate more diverse and varied contributions, suggesting that they engage with different topics and concepts compared to the H-Index discussions. Through this analysis, our contribution lies in illustrating how a focus on knowledge claims rather than sheer publication output can provide a more nuanced understanding of scientific growth. RWMD could help to identify redundant literature, hypes, and potentially disruptive innovations, offering a fresh perspective on the evaluation of scientific contributions. This approach delineates the intellectual landscape of publications through semantic similarity measures, we propose an alternative approach to addressing the challenges posed by the oversupply of publications. It not only helps recognize areas saturated with repetitive literature but also in uncovering novel ideas that could lead to significant shifts in research paradigms.

As the scientific community continues to expand, accompanied by rapid and uncontrolled growth of publications, the methodologies presented in this study could represent a foundational tool for future research assessments, encouraging a shift towards more meaningful and impactful scientific inquiry. Although it has valuable insights, this study has some limitations. Future research should focus on integrating a more diverse array of semantic analysis tools and embedding models to create a more dynamic and comprehensive framework for assessing scientific literature. The current study was limited to a comparative analysis among a set number of documents. Expanding the dataset would allow for a more comprehensive understanding of the semantic landscape across different scientific domains. In addition, in this study the GloVe was used for generating word embeddings; however, other models like Word2Vec, SciBERT, and BERT offer could be assessed for capturing contextual meanings and could potentially alter the semantic similarity outcomes.

**Open science practices**
This study embraces open science practices, reflecting our commitment to transparency and accessibility in research. Data are freely available at the following link https://zenodo.org/records/10997154, ensuring that interested parties can access and verify our findings. The analysis using R, an open-access software, aligns with our dedication to utilizing tools that promote collaboration and reproducibility in scientific research.

**Author contributions**
**Massimo Aria, Nicolás-Robinson García, Luca D'Aniello, Corrado Cuccurullo:** Conceptualization; **Massimo Aria, Nicolás-Robinson García, Luca D'Aniello** Methodology; **Luca D'Aniello.**: Data curation, Writing- Original draft preparation; **Massimo Aria, Nicolás-Robinson García**: Supervision; **Massimo Aria, Nicolás-Robinson García, Luca D'Aniello, Corrado Cuccurullo:** Writing- Reviewing and Editing

**Competing interests**
The authors declare that they have no competing interests.

**Funding information**
This research has been financed by the following research projects: PRIN-2022 "SCIK-HEALTH" (Project Code: 2022825Y5E; CUP: E53D2300611006); PRIN-2022 PNRR "The value of scientific production for patient care in Academic Health Science Centres" (Project Code: P2022RF38Y; CUP: E53D23016650001).